\documentstyle[aps,prd,preprint]{revtex}
\input epsf
\tightenlines

\newcommand{\beq}{\begin{equation}}
\newcommand{\eeq}{\end{equation}}

\begin{document}

\title{Eternal inflation and chaotic terminology}

\author{Alexander Vilenkin}

\address{ Institute of Cosmology, Department of Physics and Astronomy,\\ 
Tufts University, Medford, MA 02155, USA}

\maketitle

\begin{abstract}

``Eternal inflation'' is often confused with ``chaotic
inflation''. Moreover, the term ``chaotic inflation'' is being used in
three different meanings (all of them unrelated to eternal
inflation). I make some suggestions in an effort to untangle this
terminological mess. I also give a brief review of the origins of
eternal inflation.

\end{abstract}

\section{Introduction}

The idea of eternal inflation was introduced more than 20 years
ago. For most of this time it remained an esoteric subject, pursued by
a few enthusiasts and largely ignored by the rest of the physics
community. But in the last few years a new ``multiverse'' cosmological
paradigm has emerged with eternal inflation at its core. This led to
proliferation of papers on the subject, both in technical journals and
in the popular science literature. Unfortunately, many of these papers
(see, e.g., \cite{Rees,Banks,Tegmark,Ellis}) confuse eternal
inflation with another important idea, or rather two ideas, known
under the name of ``chaotic inflation''.

I am writing this brief note in an effort to disentangle these
different and largely unrelated concepts. The next section reviews the
origins of the key ideas of eternal inflation. (As almost everything
in eternal inflation, this has also become a somewhat controversial
subject.) Section 3 explains the different meanings of ``chaotic
inflation''. Section 4 reviews the cases where inflation is both
``chaotic'' and eternal. Finally, in Section 5 I propose some modest
measures that might help to restore clarity. I also briefly delve in
psychology and discuss possible causes for the mixup.

\section{A brief history of eternal inflation}

The eternal character of inflation was encountered in Guth's 1981
paper \cite{Guth}, where he introduced the idea of inflation. Guth
realized that the horizon and flatness problems of standard cosmology
can be solved by inflation -- a period of accelerated expansion driven
by false vacuum energy. He discovered, however, that inflation was not
easy to end. Metastable false vacuum decays through bubble
nucleation. Bubbles expand at speeds approaching the speed of light,
but the false vacuum regions that separate them expand even
faster. Thus, the bubbles never fill the entire space and inflation
never ends. This is the graceful exit problem of the old inflationary
scenario.

The problem was solved with the advent of ``new'' inflation
\cite{Linde,Albrecht}. This type of models assumes a scalar field -- 
the inflaton -- with a slowly varying potential. The field starts
somewhere near the top of the potential and slowly rolls downhill. In
the meantime, its potential energy drives the inflationary expansion
of the universe. 

The early versions of this scenario postulated a small barrier near
the top of the potential -- a leftover from the ``old'' model of
inflation. With this additional feature, the model combines the
characteristics of ``old'' and ``new'' inflation. It has a metastable
false vacuum, which decays through bubble nucleation. The slow roll of
the field then occurs inside the bubbles. Steinhardt \cite{Steinhardt}
pointed out that the global structure of the universe in this type of
model is the same as in ``old'' inflation: the bubbles are driven
apart by the inflating false vacuum and never fill the space, so
inflation never ends in the entire universe. A narrow barrier on top
of a smooth energy hill is a rather unnatural feature, and once it
was recognized that this feature is unnecessary, it was quickly
abandoned \cite{foot1}.

Eternal inflation as a generic phenomenon, which is common to a very
wide class of models, was first discussed in my paper \cite{AV83} (see
also \cite{AV84}). The evolution of the inflaton field is influenced
by quantum fluctuations, which can be pictured as a random walk of the
field with a step $\Delta\phi\sim H/2\pi$ on a horizon scale ($l\sim
H^{-1}$) per expansion time ($\Delta t\sim H^{-1}$). (Here, $H$ is the
expansion rate of the universe.) In models of ``new'' inflation,
quantum fluctuations represent a small perturbation to the classical
motion of the field, except near the maximum of the potential, where
the classical driving force is small and the quantum random walk
dominates the dynamics. The key point is that when the field is
hovering near the top of the potential, the corresponding regions
exponentially multiply. In some of these regions the field ``walks''
to the steeper part of the slope and starts rolling down, but this
``decay'' of inflating regions is much slower than their
``reproduction''. As a result, the total volume occupied by inflating
regions grows (exponentially) with time.

Eternal inflation provides a natural arena for the ``anthropic
principle''. Different thermalized regions of the eternally inflating
universe can be characterized by different values of the low-energy
constants of Nature. The values we observe here can then be largely
determined by anthropic selection.  This was first emphasized by Linde
(see \cite{Lindebook} and refernces therein).

Mathematically, an eternally inflating universe can be described by a
distribution $P(\phi,t)d\phi$, which gives the fraction of the
comoving volume of the universe where the inflaton field $\phi$ is in
the interval $d\phi$ at time $t$. This distribution satisfies a
Fokker-Planck equation. Its aproximate form for a flat potential was
given in \cite{AV83}, and the general form was derived by Starobinsky
\cite{Starobinsky}. This formalism was later extended from the 
comoving to the physical volume distribution \cite{GLM}; for a review
see \cite{LLM}.

The Fokker-Planck equation has also been extended to include the variable
``constants'' of Nature, and the resulting distributions have been used to
assign probabilities to different values of the constants.  The
question of how observational predictions are to be extracted from this
formalism is still being debated. The main difficulty is that the
distributions are sensitive to how one defines the time variable $t$
\cite{LLM,GBLL,LLM'}, and no particular choice appears to be
preferred. Possible ways to resolve this problem have been proposed in
\cite{AV98,VVW,GV'}. 

Although inflation is eternal to the future, inflationary spacetimes
are necessarily past-incomplete (see \cite{BGV} and references
therein). Hence, inflation must have some sort of a beginning. (This
issue is somewhat related to the topic of ``eternal chaotic
inflation'', to be discussed below.)

\section{The three meanings of chaotic inflation}

The term ``chaotic inflation'' first appeared in Linde's 1983 paper
\cite{Linde83}, where he proposed two different ideas. First, he realized 
that inflation is posible with potentials $V(\phi)$ which grow
unboundedly with $\phi$ and do not have a maximum. The simplest
example is a power-law potential, $V(\phi)\propto \phi^n$. Linde
showed that in order to have inflation in this type of model, the
field has to start its slow roll at a large value, $\phi \gtrsim M_p$,
where $M_p$ is the Planck mass.

As a motivation for the large initial value of the field, Linde
suggested that the universe might have started in a chaotic initial
state, with the inflaton field varying wildly from one place to
another. Inflation will then occur in regions where the field happend
to be large. It should be emphasized, however, that such a chaotic
beginning is not mandatory for inflation with an unbounded
potential. One alternative is provided by quantum cosmology: small
closed universes can spontaneously nucleate with all possible values
of the field. The corresponding probability distribution depends on
the choice of the boundary conditions for the wave function of the
universe (see \cite{debate} for a critical review), but regardless of
the choice, there is always a nonzero probability for $\phi>M_p$.

The term ``chaotic inflation'' is now being used in reference to
inflation with a chaotic beginning, with an unbounded potential, or
with both of the above. This is, of course, confusing, and the
confusion is exacerbated by the mixup with eternal inflation.

\section{Eternal chaotic inflation}

In 1986 Linde showed that inflation with an unbounded potential is
also generically eternal \cite{Linde86}\footnote{The term ``eternal
inflation'' was introduced in this paper.}. This is somewhat
unexpected, because, for a power-law potential the slope of the
potential grows with $\phi$. However, the step of the random walk
$\Delta\phi \sim H/2\pi$ is a growing function of $\phi$ as well, and
Linde found that quantum fluctuations dominate over classical dynamics
at sufficiently large $\phi$. Hence, inflation is eternal, by the same
argument as before.

Eternal inflation with unbounded potentials has some qualitatively new
features. The higher the field gets up the slope of the potential, the
faster is the rate of inflationary ``reproduction'', and the larger
are the quantum jumps of the field. As a result, in some regions the
field exhibits a runaway behavior and is driven all the way up to the
Planck density.

This scenario is sometimes called ``eternal chaotic inflation''. The
term is rather ambiguous, because of the multiple meanings of
``chaotic''.

\section{A proposal}

It would be interesting to figure out why eternal and chaotic
inflations got so much confused in the first place. One obvious reason
is that they are associated with the same physicist who made important
contributions to both subjects. Another reason, I think, is the
stochastic character of eternal inflation.  The inflating regions form
a self-similar fractal \cite{Mukunda,Winitzki}, and computer
simulations of eternal inflation \cite{Mukunda,LLM,VVW} present a
rather ``chaotic'' appearance. An even more haphazard picture, with
episodes of inflation recurring within already thermalized regions,
resulting in new islands of eternal inflation, was presented in
\cite{GV}.

History aside, chaotic usage of terms has reached epidemic
proportions. I, therefore, humbly suggest that the reader considers
taking the following preventive steps.

1. If you write about eternal inflation, refrain from calling it
``chaotic''.

2. To differentiate between the different meanings of ``chaotic
inflation'', I suggest using one of the following: (i) ``inflation
with an unbounded potential'' (or, if you like, ``topless
inflation''), (ii) ``inflation with a chaotic beginning'', or, if
necesary, (iii) ``inflation with an unbounded potential and a chaotic
beginning''.

3. If you want to refer to eternal inflation which is also chaotic, in
any of the three meanings of the word, I suggest that you follow the
same strategy as in 2 above and say, for example, ``eternal inflation
with an unbounded potential'', etc.

I am grateful to Jaume Garriga and Ken Olum for comments on the draft.


\begin{thebibliography}{99}

\bibitem{Rees}
M.J. Rees, {\it Before the Beginning}, Addison-Wesley, Reading, 1997.

\bibitem{Banks}
T. Banks and W. Fischler, astro-ph/0307459.

\bibitem{Tegmark}
M. Tegmark, astro-ph/0302131.

\bibitem{Ellis}
G.F.R. Ellis, U. Kirchner and W.R. Stoeger, astro-ph/0305292; 
W.R. Stoeger, G.F.R. Ellis and U. Kirchner, astro-ph/0407329.

\bibitem{Guth}
A.H. Guth, Phys. Rev. {\bf D23}, 347 (1981).

\bibitem{Linde}
A.D. Linde, Phys. Lett. {\bf 108B}, 389 (1982).

\bibitem{Albrecht}
A. Albrecht and P.J. Steinhardt, Phys. Rev. Lett. {\bf 48}, 1220 (1982).

\bibitem{Steinhardt}
P.J. Steinhardt, in {\it The Very Early Universe}, ed. by
G.W. Gibbons, S.W. Hawking and S.T.C. Siklos (Cambridge University
Press, 1983).

\bibitem{foot1}
Inflaton potentials with a barrier reappeared briefly in the context
of ``open inflation'' models in mid-1990's.

\bibitem{AV83}
A. Vilenkin, Phys. Rev. {\bf D27}, 2848 (1983).

\bibitem{AV84}
A. Vilenkin, Tufts preprint TUTP-84-5 (1984).

\bibitem{Lindebook}
A.D. Linde, {\it Particle Physics and Inflationary Cosmology}, Harwood
Academic, Chur, 1990.

\bibitem{Starobinsky}
A.A. Starobinsky, in {\it Lecture Notes in Physics}, vol. 246
(Springer, Heidelberg, 1986).

\bibitem{GLM}
A.S. Goncharov, A.D. Linde and V.F. Mukhanov, Int. J. Mod. Phys. {\bf
A2}, 561 (1987).

\bibitem{LLM}
A.D. Linde, D.A. Linde and A. Mezhlumian, Phys. Rev. {\bf D49}, 1783
(1994).

%\bibitem{LM}
%A.D. Linde and A. Mezhlumian, Phys. Lett. {\bf B307}, 25 (1993).

\bibitem{GBLL}
J. Garcia-Bellido, A.D. Linde and D.A. Linde, Phys. Rev. {\bf D50},
730 (1994).

\bibitem{LLM'}
A.D. Linde, D.A. Linde and A. Mezhlumian, Phys. Lett. {\bf B345}, 203
(1995).

\bibitem{AV98}
A. Vilenkin, Phys. Rev. Lett. {\bf 81}, 5501 (1998).

\bibitem{VVW}
V. Vanchurin, A. Vilenkin and S. Winitzki, Phys. Rev. {\bf D61},
083507 (2000).

\bibitem{GV'}
J. Garriga and A. Vilenkin, Phys. Rev. {\bf D64}, 023507 (2001).

\bibitem{BGV}
A. Borde, A.H. Guth and A. Vilenkin, Phys. Rev. Lett. {\bf 90}, 151301
(2003).

\bibitem{Linde83}
A.D. Linde, Phys. Lett. {\bf B129}, 177 (1983).

\bibitem{debate}
A. Vilenkin, gr-qc/9812027.

\bibitem{Linde86}
A.D. Linde, Phys. Lett. {\bf B175}, 395 (1986).

\bibitem{Mukunda}
M. Aryal and A. Vilenkin, Phys. Lett. {\bf B199}, 351 (1987).

\bibitem{Winitzki}
S. Winitzki, Phys. Rev. {\bf D65}, 083506 (2002).



\bibitem{GV}
J. Garriga and A. Vilenkin, Phys. Rev. {\bf D57}, 2230 (1988).

\end{thebibliography}
\end{document}